\documentclass{article}
\usepackage{amsmath,graphicx}
\usepackage[preprint]{spconf}
\copyrightnotice{979-8-3503-0689-7/23/\$31.00~\copyright2023 IEEE}
\toappear{To appear in {\it Proc.\ ASRU 2023}}

\usepackage{amsfonts}
\usepackage{booktabs}
\usepackage{setspace}

\usepackage{cite}
\usepackage{subcaption}
\usepackage{url}

\title{DIFFUSION-BASED MEL-SPECTROGRAM ENHANCEMENT FOR PERSONALIZED SPEECH SYNTHESIS WITH FOUND DATA}

\name{Yusheng Tian,  Wei Liu, Tan Lee}
\address{Department of Electronic Engineering, The Chinese University of Hong Kong, Hong Kong SAR}

\begin{document}


%
\maketitle

\begin{abstract}
\begin{spacing}{0.97}
Creating synthetic voices with found data is challenging, as real-world recordings often contain various types of audio degradation. One way to address this problem is to pre-enhance the speech with an enhancement model and then use the enhanced data for text-to-speech (TTS) model training. This paper investigates the use of conditional diffusion models for generalized speech enhancement, which aims at addressing multiple types of audio degradation simultaneously. The enhancement is performed on the log Mel-spectrogram domain to align with the TTS training objective. Text information is introduced as an additional condition to improve the model robustness. Experiments on real-world recordings demonstrate that the synthetic voice built on data enhanced by the proposed model produces higher-quality synthetic speech, compared to those trained on data enhanced by strong baselines. Code and checkpoints of the proposed enhancement model are available at \texttt{https://github.com/dmse4tts/DMSE4TTS}.
\end{spacing}
\end{abstract}
\begin{spacing}{0.97}

\section{Introduction}
\label{sec:intro}

Text-to-speech (TTS) models are typically trained using carefully recorded databases. Collecting such recordings is costly and sometimes impractical. In real-world applications, sometimes only low-quality recordings from the target speaker(s) are available. This has motivated the study of speech synthesis with found data, i.e., speech data that are not purposely recorded for the development of TTS systems\cite{valentini2018tasplp,valentini2016speech,wgan, improved_segan, founddata,commonvoice,hsu2019disentangling, zhang2021denoispeech}.

Developing TTS systems in this context is challenging, given the highly varied degradation of audio quality affecting the found data. Synthetic voices built directly with low-quality recordings would inevitably produce distorted speech. Previous studies have attempted to tackle this problem within the TTS framework, by augmenting the acoustic model with an additional noise embedding \cite{hsu2019disentangling, zhang2021denoispeech}. This enabled TTS model training with noisy speech, but did not consider the impact of other forms of audio degradation. Background noise is not the only disturbance present in real-world recordings. With this in mind, it might be more practical to pre-enhance found data with a separate generalized speech enhancement model \cite{hifiganSE, cascaded, generalized, voicefixer, universal, miipher}, which aims at addressing multiple types of audio degradation simultaneously. 

In the present study, we propose to use conditional diffusion models for generalized speech enhancement, and apply enhancement directly to log Mel-spectrograms to align with the TTS training objective. The choice of a diffusion model is motivated by the work of Palette \cite{saharia2022palette}, which used a single generalist diffusion model to deal with a range of image-to-image translation tasks. Mel-spectrograms are time-frequency representations that can be treated as images. Thus we expect that the diffusion model would be effective in Mel-spectrogram enhancement. To improve the model robustness against unseen forms of audio degradation, text content of speech is used as an additional condition, which is typically available in the context of TTS training.

We applied the proposed enhancement model to a real case of speech generation: developing a personalized synthetic voice for a male Cantonese speaker with 37-minute found recordings. This gentleman lost the ability to speak after receiving laryngectomy a few years ago. The recordings he provided, though containing multiple types of degradation, are the only available and precious record of his voice. TTS models are trained with speech enhanced by different systems. Subjective evaluation by human listeners show that the synthetic voice built on speech enhanced by the proposed model is rated higher for both cleanliness and overall impression, compared to those trained on data enhanced by strong speech enhancement baselines.

\section{Related Work}
\vspace{-0.2em}
\subsection{Speech synthesis with found data} 
A number of previous studies have approached the problem of speech synthesis with found data. \cite{commonvoice, founddata} designed algorithms to automatically select clean recordings from crowd-sourced data. Others consider the situation when there are no high-quality samples in the found data. \cite{hsu2019disentangling, zhang2021denoispeech} augmented the TTS model with an additional noise embedding, which can separate environmental noise from clean speech during TTS model training. Another line of research \cite{valentini2018tasplp,valentini2016speech,wgan, improved_segan} approached the problem in two steps: pre-enhance the low-quality speech audio and then use the enhanced data for TTS model training.

\subsection{Generalized speech enhancement}
Several prior works \cite{hifiganSE, cascaded, generalized, voicefixer, universal} have approached the problem of generalized speech enhancement, i.e., addressing multiple types of audio degradation simultaneously. Their successes rely on well designed simulation of audio degradation, and advanced neural network architectures. Most of these models operate on the waveform or magnitude spectrogram. The enhanced speech might be suboptimal for training TTS acoustic models, which are typically designed to predict compact acoustic representations such as Mel-spectrograms. In the recent work of Miipher \cite{miipher}, self-supervised speech representations extracted from w2v-BERT are used as the features for generalized speech enhancement. To account for the loss of speaker information in w2v-BERT features, an additional speaker embedding network is required.

\vspace{-0.3em}

\subsection{Diffusion-based speech enhancement}
The use of diffusion models for speech enhancement has been investigated in \cite{universal,restore,lu2021study, cdiff}. Most of these models are derived from diffusion-based neural vocoder\cite{diffwave}, by replacing the clean Mel-spectrogram input with a degraded one. This design may not have fully exploited the potential of diffusion models for speech enhancement, as it also incorporates the task of speech waveform generation. Our work is most similar to that of \cite{stft1}, in which diffusion models are used for speech enhancement in the complex short-time Fourier transform (STFT) domain. However, their model targets only at denoising, and is not tailored for developing TTS systems.

\section{Diffusion-based Mel-spectrogram Enhancement}
Suppose we are given a large collection of Mel-spectrogram pairs, denoted as \(\mathcal{D}=\{\boldsymbol{x}^{(i)},\boldsymbol{y}^{(i)}\}_{i=1}^N\), where \(\boldsymbol{x}^{(i)}\) represents the Mel-spectrogram of a high-quality speech sample,  and \(\boldsymbol{y}^{(i)}\) represents the Mel-spectrogram of a respective degraded sample. \(\boldsymbol{y}^{(i)}\) can be created by applying artificial audio degradation to \(\boldsymbol{x}^{(i)}\). We are interested in learning the conditional distribution \(P(\boldsymbol{x}|\boldsymbol{y})\) on \(\mathcal{D}\). If \(\mathcal{D}\) is constructed to be representative, Mel-spectrogram enhancement can be achieved by sampling from the learned conditional distribution. Diffusion models are adopted here to learn a parametric approximation of \(P(\boldsymbol{x}|\boldsymbol{y})\).

\vspace{-0.2em}
\subsection{Conditional diffusion process}
We consider the Variance Preserving (VP) diffusion model \cite{song2020score,ho2020denoising}. Suppose \(\boldsymbol{x}_0 \sim P(\boldsymbol{x}|\boldsymbol{y})\) is one enhanced realization for the degraded input \(\boldsymbol{y}\). VP diffusion defines the forward process as
\begin{align}
\mathrm{d}\boldsymbol{x}_t =-\frac{1}{2}\beta_t\boldsymbol{x}_t\mathrm{d}t+\sqrt{\beta_t}\mathrm{d}\boldsymbol{w}_t\ ,
\label{equation:forward}
\end{align}
%
where \(t\sim \mathcal{U}(0,1)\), \(\beta_t=\beta_0+\beta_1t\) is a predefined linear noise scheduler, and \(\boldsymbol{w}_t\) is a standard Brown motion.

One important result derived from (\ref{equation:forward}) is the conditional distribution of \(\boldsymbol{x}_t\) given \(\boldsymbol{x}_0\):
\begin{align}
P(\boldsymbol{x}_t|\boldsymbol{x}_0) = \mathcal{N}(\boldsymbol{\rho}(\boldsymbol{x}_0, t), \sigma^2_t\boldsymbol{I})\ ,
\label{equation:condition}
\end{align}
where \(\boldsymbol{\rho}(\boldsymbol{x}_0, t)=e^{-\frac{1}{2}\int_0^t\beta_s \mathrm{d}s}\boldsymbol{x}_0\), and \(\sigma_t^2=1-e^{-\int_0^t \beta_s \mathrm{d}s}\). The result given by (\ref{equation:condition}) suggests that if \(\boldsymbol{x}_0\) is known, we can sample \(\boldsymbol{x}_t\) using the reparameterization trick:
\begin{align}
\boldsymbol{x}_t = \boldsymbol{\rho}(\boldsymbol{x}_0, t) + \sigma_t \boldsymbol{\epsilon}_t,\boldsymbol{\epsilon}_t \sim \mathcal{N}(\boldsymbol{0}, \boldsymbol{I})\ .
\label{equation:forward_sim}
\end{align}
Furthermore, as \(t\xrightarrow{}1\), with appropriate noise scheduler \(\beta_t\) we have \(\boldsymbol{\rho}(\boldsymbol{x}_0, t)\xrightarrow{}\boldsymbol{0}\) and \(\sigma_t\xrightarrow{}1\), meaning that the forward process gradually transforms the data distribution from \(P(\boldsymbol{x}|\boldsymbol{y})\) into a standard Gaussian distribution \(\mathcal{N}(\boldsymbol{0}, \boldsymbol{I})\). 

Diffusion models generate samples by reversing the above forward process, starting with a Gaussian noise:
\begin{align}
\mathrm{d}\boldsymbol{x}_t = -\frac{1}{2}\beta_t\left[\boldsymbol{x}_t +\nabla_{\boldsymbol{x}_t}\log P(\boldsymbol{x}_t|\boldsymbol{y})\right]\mathrm{d}t \ .
\label{equation:reverse}
\end{align}
Note that the reverse process is conditioned on \(\boldsymbol{y}\) to enable conditional generation. The core part of a diffusion model is to train a neural network \(S_{\boldsymbol{\theta}}\) to estimate the value of \(\nabla_{\boldsymbol{x}_t}\log P(\boldsymbol{x}_t|\boldsymbol{y})\) (a.k.a the score). Once the score is known for all time steps, we can draw samples from \(P(\boldsymbol{x}|\boldsymbol{y})\) by simulating the reverse process from \(t=1\) to \(0\), for example with an ODE solver\cite{lu2022dpmsolver}.
\vspace{-0.7em}

\subsection{Robust text condition}
\vspace{-0.2em}
As mentioned earlier, the Mel-spectrogram enhancement model relies on a synthetic dataset of paired samples to learn the conditional distribution \(P(\boldsymbol{x}|\boldsymbol{y})\). Therefore a domain gap between training samples and real-world degraded recordings is inevitable. Consequently, the trained model may overfit to in-domain data and fail to generalize well to unseen forms of audio degradation.

In order to improve the model robustness, we introduce text content of speech samples as an additional condition. Text transcription is usually available in the context of TTS development, and has shown to improve the robustness of speech enhancement models\cite{text1, text2}. Inspired by GradTTS \cite{popov2021grad}, a diffusion-based TTS model, we use an average Mel-spectrogram \(\boldsymbol{\mu}\) to represent text. \(\boldsymbol{\mu}\) is of the same shape as \(\boldsymbol{y}\) and is obtained in three steps. First, text transcription for each training sample is converted to time-aligned phone sequence by forced alignment. Second, a phoneme-to-Mel-spectrum dictionary is created on training data by averaging speech frames that correspond to the same phoneme. Then given any time-aligned phone sequence, the respective average Mel-spectrogram \(\boldsymbol{\mu}\) is obtained by looking up the dictionary. 

When text is provided, the reverse process is rewritten as:
\begin{align}
\mathrm{d}\boldsymbol{x}_t = -\frac{1}{2}\beta_t\left[\boldsymbol{x}_t +\nabla\log P(\boldsymbol{x}_t|\boldsymbol{y}, \boldsymbol{\mu})\right]\mathrm{d}t \ .
\label{equation:reversemu}
\end{align}

\subsection{Training and inference}
\begin{figure}[t]
  \centering
  \includegraphics[width=0.9\linewidth]{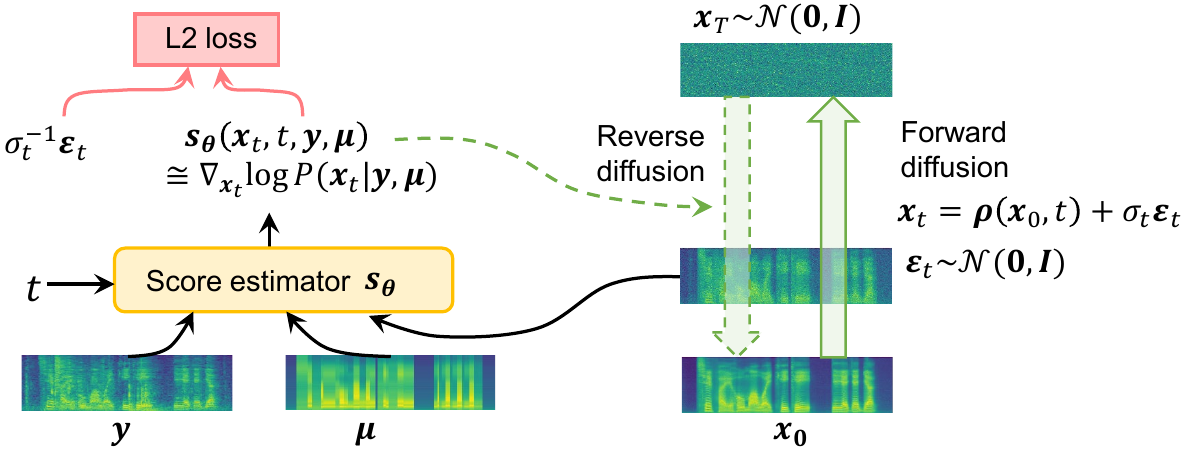}
  \caption{The training (solid lines) and inference (dashed lines) procedure of the proposed enhancement model.}
  \label{fig:system_design}
  \vspace{-1.2em}
\end{figure}

Following \cite{song2020score}, we make the score estimator \(S_{\boldsymbol{\theta}}\) aware of the time step and train it with a weighted \(L2\) loss:
\begin{equation}
\mathcal{L}(\boldsymbol{\theta})=\mathbb{E}_{t} \sigma_t^2{\mathbb{E}}_{(\boldsymbol{x}_0, \boldsymbol{y})} {\mathbb{E}}_{\boldsymbol{\epsilon}_t} \left\Vert S_{\boldsymbol{\theta}}(\boldsymbol{x}_t, t, \boldsymbol{y},\boldsymbol{\mu}) +\sigma_t^{-1}\boldsymbol{\epsilon}_t\right\Vert_2^2\ ,
    \label{equation:loss}
\end{equation}
where we have made use of the following results:
\begin{align}
P(\boldsymbol{x}_t|\boldsymbol{x}_0,\boldsymbol{y}, \boldsymbol{\mu}) = P(\boldsymbol{x}_t|\boldsymbol{x}_0) =\mathcal{N}(\boldsymbol{\rho}(\boldsymbol{x}_0, t), \sigma^2_t\boldsymbol{I})\ ,
\label{equation:indp}
\end{align}
\vspace{-2.2em}
\begin{align}
\nabla_{\boldsymbol{x}_t} \log P(\boldsymbol{x}_t|\boldsymbol{x}_0,\boldsymbol{y}, \boldsymbol{\mu}) = -\sigma_t^{-1}\boldsymbol{\epsilon}_t\ .
\label{equation:gauss}
\end{align}
Once the score estimator is trained, we can use the predicted score to generate samples by running equation (\ref{equation:reversemu}) backward in time from \(t=1\) to \(0\). The training and inference procedures of the proposed Mel-spectrogram enhancement model are illustrated as in Figure \ref{fig:system_design}.
\end{spacing}

\section{Experimental Setup}
\begin{spacing}{0.97}
\subsection{Baselines for comparison}
\vspace{-0.2em}
The proposed enhancement model is named as DMSEtext, abbreviating \textbf{D}iffusion-based \textbf{M}el-\textbf{S}pectrogram \textbf{E}nhancement with \textbf{text} conditioning. The model without text conditioning is named as DMSEbase. We compare them with two baselines: Demucs\cite{Demucus} and VoiceFixer\cite{voicefixer}. Demucs is a denoising model. We use it as a baseline to investigate the efficacy of a single-task denoising model in the intended application. VoiceFixer is a regression model trained to address a range of audio degradation, including additive noise, reverberation, clipping and low-bandwidth. The motivation of choosing VoiceFixer as a baseline is to examine whether diffusion-based speech enhancement has advantage over a regression-based approach.

\subsection{Network architecture}
 The GradTTS model architecture\cite{popov2021grad} is adopted for the proposed Mel-spectrogram enhancement model. It is augmented to be conditioned on both text (\(\boldsymbol{\mu}\)) and audio \((\boldsymbol{y}\)). Conditions are provided to the score estimator by concatenation in the channel dimension. We scale the model depth to 5-layer, with the output channel dimensions being 32, 64, 128, 256, 256 respectively. The TTS model is similar to DurIAN\cite{durian}, except that style control and postnet are not included. We use the HiFi-GAN\cite{hifigan} neural vocoder to convert log Mel-spectrograms into waveforms. The number of Mel bands is increased from 80 to 128. We empirically found that this is beneficial for modelling voices of low vocal range, as is the case of our target speaker.
 \vspace{-0.5em}
 
 \subsection{Data}
 The found data from our target speaker contain 513 utterances of spontaneous yet very fluent speech, giving a total duration of approximately 37 minutes, all manually transcribed. Speech was recorded in six different sessions with varied room acoustics and sampled at 22.05 kHz. Audio degradation types found in the recordings include background noise, room reverberation, band limiting and magnitude clipping. The speech is in Cantonese. This speech corpus is used for performance evaluation of speech enhancement models. The enhanced speech is for training the TTS model.
 
 We use CUSENT\cite{lee2002spoken}, a multi-speaker Cantonese speech database and artificially created degraded audio to train DMSEbase and DMSEtext. CUSENT contains about 20 hours read speech from 80 speakers, sampled at 16kHz. It was originally designed for automatic speech recognition (ASR) and the recordings contain low-level noise, making it less than ideal for training speech enhancement models . We therefore run Demucs on CUSENT to obtain high-quality clean speech. The resulted dataset, denoted by DenoiseCUSENT, provides the clean reference for speech enhancement training. Data from speaker cn01m and cn12f are held out for validation.
 
 To synthesize distorted speech, we consider the four most frequent types of audio degradation present in found data: noise, reverberation, band limiting and magnitude clipping. We use the DNS noise dataset\cite{dnsnoise} and the RIRs\cite{rirnoise} dataset to simulate background noise and reverberation respectively, adopting their default train/test split. Band limiting and magnitude clipping are simulated with the Scipy signal processing toolkit\footnote{\url{https://docs.scipy.org/doc/scipy/reference/signal.html}}. The four types of degradation are applied following a specific order: reverberation, noise, clipping and band limiting. Parameters such as SNR or frequency are randomly set to be within a reasonable range. Details of simulated degradation are released in the project repository.

For fair comparison with Demucs and VoiceFixer, which are trained on English datasets, we also train a diffusion-based enhancement model on the high-quality English dataset VCTK\cite{VCTK}, to rule out the impact of language mismatch. Data from speaker p232 and p257 are held out from training, as they appear in a standard denoising testset. Text conditioning is not included here as the language of found data is Cantonese. This model is referred to as DMSEbase(VCTK).

Throughout the experiments Mel-spectrograms are computed on audio signals resampled to 22.05 kHz with a window length of 1024 and a hop size of 256. The number of mel filterbanks is set to 128 as mentioned earlier.
 
\subsection{Implementation details}
DMSEbase and DMSEtext are trained on DenoiseCUSENT for 900 epochs, while DMSEbase(VCTK) is trained on VCTK for 750 epochs, all at a batch size of 32. The number of reverse steps for all three models is 25 with the DPMsolver \cite{lu2022dpmsolver}. Mel-spectrograms are mean-normalized and scaled to be within the range of \([-1.0, 1.0]\). We empirically found that this normalization helps improving the convergence speed of reverse diffusion. 

The TTS model is pre-trained on CUSENT for 600 epochs with a batch size of 32, then fine-tuned on the target speaker's data (original or enhanced) for 6000 steps with a batch size of 16. The neural vocoder is fine-tuned from a pre-trained HiFi-GAN\footnote{\url{https://github.com/jik876/hifi-gan}} on the VCTK and CUSENT dataset for 220,000 steps with a batch size of 32. We insert a pre-processing block to map the 128-dim log Mel-spectrogram input to 80-dim, to take advantage of the pre-trained HiFi-GAN parameters. 

We use the Adam optimizer\cite{Adam} in all experiments with the default value \([0.9, 0.999]\) for betas. Learning rate is fixed at 1e-4, 1e-3 and 1e-6 for training the DMSEbase/base(VCTK)/text, the TTS model, and the HiFi-GAN vocoder respectively.

Time-aligned phone sequences for training the TTS model and DMSEtext are obtained using the Montreal Forced Aligner\cite{mfa}. We apply denoising on found data with Demucs before running forced alignment, in order to improve the alignment accuracy. Note that the denoised data are used only during forced-alignment. All enhancement models take the original found data as input for evaluation.

We additionally train a CTC-based ASR model for objective evaluation of speech enhancement, following the recipe from SpeechBrain\footnote{\url{https://github.com/speechbrain/speechbrain/tree/develop/recipes/TIMIT/ASR/CTC}}. The model is trained on KingASR086, a commercial Cantonese speech recognition corpus purchased from SpeechOcean\footnote{\url{https://en.speechocean.com/datacenter/recognition.html}}, which contains 80-hour reasonably clean read speech from 136 speakers, sampled at 44.1 kHz. No data augmentation is applied.

\section{Result}
\begin{figure*}[t]
     \centering
     \begin{subfigure}[b]{0.95\textwidth}
         \centering
         \includegraphics[width=\textwidth]{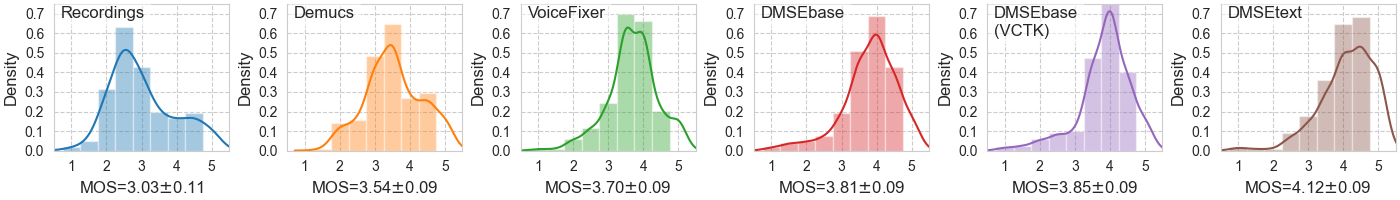}
         \caption{The distribution of MOS results for cleanliness. Mean and 95\% confidence interval are reported at the bottom.}
         \label{fig:y equals x}
     \end{subfigure}
     \begin{subfigure}[b]{0.95\textwidth}
         \centering
         \includegraphics[width=\textwidth]{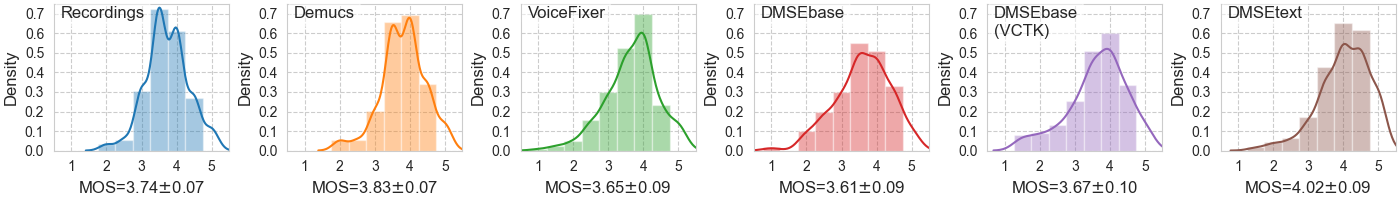}
         \caption{The distribution of MOS results for overall impression. Mean and 95\% confidence interval are reported at the bottom.}
         \vspace{-0.4em}
         \label{fig:three sin x}
     \end{subfigure}
        \caption{Score distribution of speech enhanced by different systems.}
        \label{fig:three graphs}
\end{figure*}

\begin{figure*}[t]
     \centering
\includegraphics[width=0.95\textwidth]{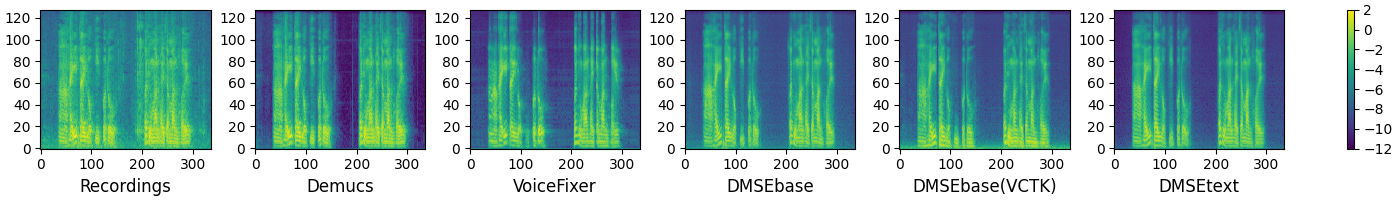}
        \caption{Comparing the log Mel-spectrograms of speech audio generated by different speech enhancement models.}
        \label{fig:melspec}
        \vspace{-1.2em}
\end{figure*}

\subsection{Enhancing found speech data}
The 513 utterances in the original found data, denoted as Recordings, are enhanced by the three diffusion-based models as well as the two baselines, respectively.

Objective evaluation was carried out with the ASR model. Phone error rate (PER\%) results are summarized in Table \ref{tab:per}. It is noted that VoiceFixer and DMSEbase/base(VCTK) tend to introduce pronunciation distortions (reflected in the number of substitutions) and erase speech segments (reflected in the number of deletions) compared with Demucs. On the other hand, text conditioning helps DMSEtext to preserve speech content and at the same time improve the audio quality over original recordings.
\begin{table}
  \caption{PER\% evaluated by a separately trained ASR model for speech recordings enhanced by different systems, as well as the original recordings. N, I, D, S stands for the number of phones in the reference text, and the number of insertions, deletions, and substitutions in decoded results, respectively.}
		\label{tab:per}
		\centering
		\renewcommand{\arraystretch}{0.95}
            \small
 		\begin{tabular}{l|c|c}
			\toprule
			Source & PER\% & Details (N / I, D, S)\\
                \midrule
			Recordings & 22.7 & 18928 / 282, 495, 3525\\
                \midrule
                Demucs &  \textbf{20.3} & 18928 / 321, 300, 3224\\
                VoiceFixer & 29.7 & 18928 / 352, 921, 4341 \\
                \midrule
                DMSEbase & 24.7 & 18928 / 317, 702, 3652 \\
                DMSEbase(VCTK) & 24.3 & 18928 / 334, 593, 3664 \\
                DMSEtext & \textbf{17.6} & 18928 / 232, 258, 2836\\
			\bottomrule
		\end{tabular}
  \vspace{-0.7em}
	\end{table}

Subjective evaluation was conducted through a web-based listening test. We selected the same 87 sentences with lengths between 4.0 and 5.0 seconds from the enhanced data produced by each system, as well as the original recordings. This gives six stimuli for each sentence. The test is evaluated in Mean Opinion Score (MOS) format. Listeners were asked to rate the cleanliness (no noise or reverberation) as well as the overall impression of each stimulus on a scale from 1.0 to 5.0. A clean audio sample from DenoiseCUSENT and a synthetic distorted sample were provided, serving as the high and low anchors. We evenly split the 87 sentences into three groups and recruited 18 listeners per group. Each listener was presented with 29 sentences, and each sentence was produced by one of the six systems, randomly selected. All listeners are native Cantonese speakers. 

The score distributions depicted in Figure \ref{fig:three graphs} indicate that VoiceFixer and DMSEbase/base(VCTK) generate speech that is clean but somewhat distorted, while Demucs gives the opposite pattern. On the other hand, DMSEtext received the highest score for both cleanliness and overall impression, which conforms with the objective evaluation results. Notably, all diffusion-based models outperformed the two baselines in terms of cleanliness. We speculate this to be related to the generative modeling approach, where the generation process is influenced by a clean speech prior.

By comparing DMSEbase against DMSEbase(VCTK), we can conclude that there is no significant gain from using a Cantonese dataset, except for the availability of text information. Figure \ref{fig:melspec} gives a specific example of the Mel-spectrograms enhanced by different systems. We can see that audio enhanced by Demucs preserves more speech content as well as disturbances than audio enhanced by other systems. 

\subsection{Speech synthesis with enhanced data}
To ensure that the model can produce high-quality speech that meets the target speaker’s personal needs, we asked him to provide text content that he would like to or most likely say in his daily life for TTS model evaluation. A total of 30 sentences were randomly selected from a script written by the target speaker, and synthesized with TTS models trained on speech enhanced by different systems. We recruited 24 native Cantonese speakers to participate in the listening test. The test format is similar to that described in the previous section. Each listener heard all 30 sentences once, and each sentence generated by one of the six TTS models was evaluated by exact four listeners. 

\begin{table}
	\caption{MOS results for synthetic voices built on data enhanced by different systems.}
		\label{tab:tts}
		\centering
		\renewcommand{\arraystretch}{0.95}
            \small
 		\begin{tabular}{l|c|c}
			\toprule
			Source & Cleanliness & Overall impression\\
                \midrule
			Recordings & 2.74 \(\pm\) 0.08 & 3.12 \(\pm\) 0.06\\
                \midrule
                Demucs &  3.22 \(\pm\) 0.08 & 3.38 \(\pm\) 0.05\\
                VoiceFixer & 3.97 \(\pm\) 0.09 & 3.95 \(\pm\) 0.09 \\
                \midrule
                DMSEbase & 4.02 \(\pm\) 0.09 & 3.98 \(\pm\) 0.08 \\
                DMSEbase(VCTK) & 4.05 \(\pm\) 0.09 & 4.02 \(\pm\) 0.09 \\
                DMSEtext & \textbf{4.32} \(\pm\) 0.08 & \textbf{4.17} \(\pm\) 0.06\\
			\bottomrule
		\end{tabular}
  \vspace{-1.2em}
	\end{table}

The MOS results in Table \ref{tab:tts} show an opposite trend to that observed in Figure \ref{fig:three graphs}, with the synthetic voice built on speech enhanced by Demucs receiving much lower scores for overall impression compared with other systems. Two factors might explain the difference. First, the remaining disturbances in speech enhanced by Demucs were turned into unpleasant averaged artifacts after TTS training. Second, the text encoder in the TTS system is shared across speakers, which might have mitigated the negative impact of content distortion for VoiceFixer and DMSEbase/base(VCTK). Nevertheless, the synthetic voice built on speech enhanced by DMSEtext received the highest score for both cleanliness and overall impression, which is expected given the speech enhancement evaluation results. Readers are encouraged to visit \texttt{https://dmse4tts.github.io/} to listen to audio samples. 
\vspace{-0.3em}

\section{Conclusion}
We introduced a diffusion-based Mel-spectrogram enhancement model, which is intended for pre-enhancing found data for TTS model training. It is designed to tackle multiple types of audio degradation simultaneously, and is conditioned on text transcriptions to improve the model robustness. We empirically showed that the use of generative modelling encourages clean speech output, and that a generalist enhancement model is preferred over a single-task denoising model for pre-enhancing real-world recordings for TTS model development. Subjective evaluations by human listeners demonstrate that the resulted synthetic voice produces higher-quality synthetic speech compared to those trained on data enhanced by strong speech enhancement baselines.
\end{spacing}

\vspace{-0.05em}
\section{Acknowledgements}
\begin{spacing}{0.97}
We would like to thank the target speaker, who generously agreed to showcase his synthetic voice on the demo page. The first author is supported by the Hong Kong Ph.D. Fellowship Scheme of the Hong Kong Research Grants Council.
\end{spacing}

\bibliographystyle{IEEEbib}
\bibliography{strings,refs}

\begin{thebibliography}{10}

\bibitem{valentini2018tasplp}
Cassia Valentini-Botinhao and Junichi Yamagishi,
\newblock ``Speech enhancement of noisy and reverberant speech for text-to-speech,''
\newblock {\em IEEE/ACM Trans. Audio Speech Lang.}, vol. 26, no. 8, pp. 1420--1433, 2018.

\bibitem{valentini2016speech}
Cassia Valentini-Botinhao, Xin Wang, Shinji Takaki, and Junichi Yamagishi,
\newblock ``Speech enhancement for a noise-robust text-to-speech synthesis system using deep recurrent neural networks.,''
\newblock in {\em Proc. INTERSPEECH}, 2016, pp. 352--356.

\bibitem{wgan}
Nagaraj Adiga, Yannis Pantazis, Vassilis Tsiaras, and Yannis Stylianou,
\newblock ``Speech enhancement for noise-robust speech synthesis using wasserstein gan.,''
\newblock in {\em Proc. INTERSPEECH}, 2019, pp. 1821--1825.

\bibitem{improved_segan}
Seyyed~Saeed Sarfjoo, Xin Wang, Gustav~Eje Henter, Jaime Lorenzo{-}Trueba, Shinji Takaki, and Junichi Yamagishi,
\newblock ``Transformation of low-quality device-recorded speech to high-quality speech using improved {SEGAN} model,''
\newblock {\em CoRR}, 2019.

\bibitem{founddata}
Pallavi Baljekar and Alan~W Black,
\newblock ``Utterance selection techniques for tts systems using found speech.,''
\newblock in {\em Proc. SSW}, 2016, pp. 184--189.

\bibitem{commonvoice}
Sewade Ogun, Vincent Colotte, and Emmanuel Vincent,
\newblock ``Can we use common voice to train a multi-speaker {TTS} system?,''
\newblock in {\em Proc. SLT}, 2022, pp. 900--905.

\bibitem{hsu2019disentangling}
Wei-Ning Hsu, Yu~Zhang, Ron~J Weiss, Yu-An Chung, Yuxuan Wang, Yonghui Wu, and James Glass,
\newblock ``Disentangling correlated speaker and noise for speech synthesis via data augmentation and adversarial factorization,''
\newblock in {\em Proc. ICASSP}, 2019, pp. 5901--5905.

\bibitem{zhang2021denoispeech}
Chen Zhang, Yi~Ren, Xu~Tan, Jinglin Liu, Kejun Zhang, Tao Qin, Sheng Zhao, and Tie-Yan Liu,
\newblock ``Denoispeech: Denoising text to speech with frame-level noise modeling,''
\newblock in {\em Proc. ICASSP}, 2021, pp. 7063--7067.

\bibitem{hifiganSE}
Jiaqi Su, Zeyu Jin, and Adam Finkelstein,
\newblock ``Hifi-gan: High-fidelity denoising and dereverberation based on speech deep features in adversarial networks,''
\newblock in {\em Proc. INTERSPEECH}, 2020, pp. 4506--4510.

\bibitem{cascaded}
Arun~Asokan Nair and Kazuhito Koishida,
\newblock ``Cascaded time+ time-frequency unet for speech enhancement: Jointly addressing clipping, codec distortions, and gaps,''
\newblock in {\em Proc. ICASSP}, 2021, pp. 7153--7157.

\bibitem{generalized}
Santiago Pascual, Joan Serr{\`{a}}, and Antonio Bonafonte,
\newblock ``Towards generalized speech enhancement with generative adversarial networks,''
\newblock in {\em Proc. INTERSPEECH}, 2019, pp. 1791--1795.

\bibitem{voicefixer}
Haohe Liu, Xubo Liu, Qiuqiang Kong, Qiao Tian, Yan Zhao, DeLiang Wang, Chuanzeng Huang, and Yuxuan Wang,
\newblock ``Voicefixer: {A} unified framework for high-fidelity speech restoration,''
\newblock in {\em Proc. INTERSPEECH}, 2022, pp. 4232--4236.

\bibitem{universal}
Joan Serr{\`{a}}, Santiago Pascual, Jordi Pons, R.~Oguz Araz, and Davide Scaini,
\newblock ``Universal speech enhancement with score-based diffusion,''
\newblock {\em CoRR}, vol. abs/2206.03065, 2022.

\bibitem{miipher}
Yuma Koizumi, Heiga Zen, Shigeki Karita, Yifan Ding, Kohei Yatabe, Nobuyuki Morioka, Yu~Zhang, Wei Han, Ankur Bapna, and Michiel Bacchiani,
\newblock ``Miipher: {A} robust speech restoration model integrating self-supervised speech and text representations,''
\newblock in {\em Proc. WASPAA}, 2023, pp. 1--5.

\bibitem{saharia2022palette}
Chitwan Saharia, William Chan, Huiwen Chang, Chris Lee, Jonathan Ho, Tim Salimans, David Fleet, and Mohammad Norouzi,
\newblock ``Palette: Image-to-image diffusion models,''
\newblock in {\em Proc. SIGGRAPH}, 2022, pp. 1--10.

\bibitem{restore}
Jianwei Zhang, Suren Jayasuriya, and Visar Berisha,
\newblock ``Restoring degraded speech via a modified diffusion model,''
\newblock in {\em Proc. INTERSPEECH}, 2021, pp. 221--225.

\bibitem{lu2021study}
Yen-Ju Lu, Yu~Tsao, and Shinji Watanabe,
\newblock ``A study on speech enhancement based on diffusion probabilistic model,''
\newblock in {\em Proc. APSIPA ASC}, 2021, pp. 659--666.

\bibitem{cdiff}
Yen-Ju Lu, Zhong-Qiu Wang, Shinji Watanabe, Alexander Richard, Cheng Yu, and Yu~Tsao,
\newblock ``Conditional diffusion probabilistic model for speech enhancement,''
\newblock in {\em Proc. ICASSP}, 2022, pp. 7402--7406.

\bibitem{diffwave}
Zhifeng Kong, Wei Ping, Jiaji Huang, Kexin Zhao, and Bryan Catanzaro,
\newblock ``Diffwave: {A} versatile diffusion model for audio synthesis,''
\newblock in {\em Proc. ICLR}, 2021.

\bibitem{stft1}
Simon Welker, Julius Richter, and Timo Gerkmann,
\newblock ``Speech enhancement with score-based generative models in the complex {STFT} domain,''
\newblock in {\em Proc. INTERSPEECH}, 2022, pp. 2928--2932.

\bibitem{song2020score}
Yang Song, Jascha Sohl{-}Dickstein, Diederik~P. Kingma, Abhishek Kumar, Stefano Ermon, and Ben Poole,
\newblock ``Score-based generative modeling through stochastic differential equations,''
\newblock in {\em Proc. ICLR}, 2021.

\bibitem{ho2020denoising}
Jonathan Ho, Ajay Jain, and Pieter Abbeel,
\newblock ``Denoising diffusion probabilistic models,''
\newblock in {\em Proc. NeurIPS}, 2020, vol.~33, pp. 6840--6851.

\bibitem{lu2022dpmsolver}
Cheng Lu, Yuhao Zhou, Fan Bao, Jianfei Chen, Chongxuan Li, and Jun Zhu,
\newblock ``{DPM}-solver: A fast {ODE} solver for diffusion probabilistic model sampling in around 10 steps,''
\newblock in {\em Proc. NeurIPS}, 2022, vol.~35, pp. 5775--5787.

\bibitem{text1}
Keisuke Kinoshita, Marc Delcroix, Atsunori Ogawa, and Tomohiro Nakatani,
\newblock ``Text-informed speech enhancement with deep neural networks,''
\newblock in {\em Proc. INTERSPEECH}, 2015, pp. 1760--1764.

\bibitem{text2}
Wei Wang, Wangyou Zhang, Shaoxiong Lin, and Yanmin Qian,
\newblock ``Text-informed knowledge distillation for robust speech enhancement and recognition,''
\newblock in {\em Proc. ISCSLP}, 2022, pp. 334--338.

\bibitem{popov2021grad}
Vadim Popov, Ivan Vovk, Vladimir Gogoryan, Tasnima Sadekova, and Mikhail Kudinov,
\newblock ``Grad-tts: A diffusion probabilistic model for text-to-speech,''
\newblock in {\em Proc. ICML}, 2021, pp. 8599--8608.

\bibitem{Demucus}
Alexandre D{\'{e}}fossez, Gabriel Synnaeve, and Yossi Adi,
\newblock ``Real time speech enhancement in the waveform domain,''
\newblock in {\em Proc. INTERSPEECH}, 2020, pp. 3291--3295.

\bibitem{durian}
Chengzhu Yu, Heng Lu, Na~Hu, Meng Yu, Chao Weng, Kun Xu, Peng Liu, Deyi Tuo, Shiyin Kang, Guangzhi Lei, Dan Su, and Dong Yu,
\newblock ``{DurIAN: Duration Informed Attention Network for Speech Synthesis},''
\newblock in {\em Proc. INTERSPEECH}, 2020, pp. 2027--2031.

\bibitem{hifigan}
Jungil Kong, Jaehyeon Kim, and Jaekyoung Bae,
\newblock ``Hifi-gan: Generative adversarial networks for efficient and high fidelity speech synthesis,''
\newblock in {\em Proc. NeurIPS}, 2020, vol.~33, pp. 17022--17033.

\bibitem{lee2002spoken}
Tan Lee, Wai~Kit Lo, Pak-Chung Ching, and Helen Meng,
\newblock ``Spoken language resources for cantonese speech processing,''
\newblock {\em Speech Commun.}, vol. 36, no. 3-4, pp. 327--342, 2002.

\bibitem{dnsnoise}
Chandan K.~A. Reddy, Vishak Gopal, Ross Cutler, Ebrahim Beyrami, Roger Cheng, Harishchandra Dubey, Sergiy Matusevych, Robert Aichner, Ashkan Aazami, Sebastian Braun, Puneet Rana, Sriram Srinivasan, and Johannes Gehrke,
\newblock ``The {INTERSPEECH} 2020 deep noise suppression challenge: Datasets, subjective testing framework, and challenge results,''
\newblock in {\em Proc. INTERSPEECH}, 2020, pp. 2492--2496.

\bibitem{rirnoise}
Tom Ko, Vijayaditya Peddinti, Daniel Povey, Michael~L. Seltzer, and Sanjeev Khudanpur,
\newblock ``A study on data augmentation of reverberant speech for robust speech recognition,''
\newblock in {\em Proc. ICASSP}, 2017, pp. 5220--5224.

\bibitem{VCTK}
Junichi Yamagishi, Christophe Veaux, and Kirsten MacDonald,
\newblock ``Cstr vctk corpus: English multi-speaker corpus for cstr voice cloning toolkit (version 0.92),''
\newblock 2019.

\bibitem{Adam}
Diederik~P. Kingma and Jimmy Ba,
\newblock ``Adam: {A} method for stochastic optimization,''
\newblock in {\em Proc. ICLR}, 2015.

\bibitem{mfa}
Michael McAuliffe, Michaela Socolof, Sarah Mihuc, Michael Wagner, and Morgan Sonderegger,
\newblock ``Montreal forced aligner: Trainable text-speech alignment using kaldi.,''
\newblock in {\em Proc. INTERSPEECH}, 2017, vol. 2017, pp. 498--502.

\end{thebibliography}

\end{document}